\documentclass[a4paper]{jpconf}

\usepackage[pdftex]{graphicx}
\usepackage{url}
\usepackage{amssymb}
\usepackage[cmex10]{amsmath}
\usepackage[figure,vlined]{algorithm2e}
\usepackage{caption}
\newcommand{\Complex}{\mathbb{C}}
\newcommand{\Real}{\mathbb{R}}
\newcommand{\hdp}{\mid}
\newcommand{\Dirac}{Dirac}
\newcommand{\tensor}{\otimes}
\newcommand{\id}[1]{I_{#1}}
\newcommand{\shift}[2]{J_{#1}^{#2}}
\newcommand{\ssum}[2]{\sum_{#1}{#2}}
\newcommand{\dsum}[2]{\bigoplus_{#1}{#2}}

\newcommand{\vol}[1]{vol(#1)}
\newcommand{\pr}[2]{P_{#1,#2}}
\newcommand{\Preconditionera}{\text{Preconditioner1}}
\newcommand{\Preconditionerb}{\text{Preconditioner2}}
\newcommand{\isInvertible}[1]{\text{invertible}(#1)}
\newcommand{\isPeriodic}[1]{\text{isPeriodic}(#1)}
\newcommand{\isDiagonal}[1]{\text{diagonal}(#1)}

\newcommand{\Print}[1]{\text{print}}
\newcommand{\Pe}{P_e}
\newcommand{\Po}{P_o}
\newcommand{\U}{U}
\newcommand{\matrixtype}[2]{#1 \times #2}
\newcommand{\type}[1]{\text{type}(#1)}

\newenvironment{template}[1]{%
\scriptsize
\begin{algorithm}
\label{#1}\small
\SetKwInOut{Constant}{Constant}
\SetKwInOut{Input}{Input}
\SetKwInOut{Output}{Output}
\SetKwInOut{Match}{Match}
\SetKwInOut{Require}{Require}
\SetKwInOut{Var}{Var}
}{%
\end{algorithm}
}
\newenvironment{goal}[1]{%
\scriptsize
\scriptsize\begin{algorithm}
\label{#1}\small
\SetKwInOut{Constant}{Constant}
\SetKwInOut{Templates}{Templates}
\SetKwInOut{Input}{Input}
\SetKwInOut{Output}{Output}
\SetKwInOut{Var}{Var}
}{%
\end{algorithm}
}
\usepackage{array}
\begin{document}
\title{Automated Code Generation for Lattice Quantum Chromodynamics and beyond}

\author{Denis Barthou}

\address{University of Bordeaux\\
LaBRI / INRIA Bordeaux Sud-Ouest}

\ead{firstname.lastname@inria.fr}

\author{Olivier Brand-Foissac,Olivier P\`ene}
\address{University of Paris Sud\\
Laboratory of Theoretical Physics}
\ead{firstname.lastname@th.u-psud.fr}

\author{Gilbert Grosdidier}
\address{
University of Paris Sud\\
Laboratoire de l'Accelerateur Lineaire}
\ead{firstname.lastname@cern.ch}

\author{Romain Dolbeau}
\address{CAPS Entreprise\\
Rennes, France }
\ead{firstname.lastname@caps-entreprise.com}

\author{Christina Eisenbeis, Michael Kruse, Konstantin Petrov (speaker)}
\address{INRIA\\
Saclay, France}
\ead{firstname.lastname@inria.fr}




\author{Claude Tadonki}
\address{Mines ParisTech\\
Fontainebleau, France }
\ead{firstname.name@mines.paristech.fr}

\begin{abstract}
We present here our ongoing work on a Domain Specific Language which aims to simplify Monte-Carlo simulations and measurements in the domain of Lattice Quantum Chromodynamics. The tool-chain, called Qiral, is used to produce high-performance OpenMP C code from LaTeX sources. We discuss conceptual issues and details of implementation and optimization. The comparison of the performance of the generated code to the well-established simulation software is also made.
\end{abstract}

\section{Introduction}
Quantum Chromodynamics (QCD) is a fundamental theory within Standard Model used to describe strong intercations inside a  nuclei. As the coupling constant for this force is large, the perturbation theory cannot be used in most cases. Instead, a number of non-perturbative methods have been developed, with the  most prominent of them being Lattice QCD (LQCD). It is formulated in discrete space-time, with the matter fields residing on sites and the interaction fields (gluons) live on links. This setup allows us to simulate it on a computer by means of Monte-Carlo simulations. Such analysis already  provided many insights into the nature of strong interactions and delivered a number of precise results to confront experimental data. With the arrival of teraflop installations, LQCD is now regarded as the most reliable method of solving QCD, as the simulations can be systematically improved. It has also been extended to help solve theories other than QCD. 

The most time-consuming part of these simulations is the frequent inversion of an immense matrix (Hopping Matrix), which encodes interactions within the simulated system. To do this efficiently, thousands of people over the globe are working for years, inventing new methods for the inversion. At the same time, new Lagrangians appear, which, while preserving physical properties, have different mathematical ones which are supposed to make Monte-Carlo simulations faster and more reliable. 
 
The Hopping Matrix is actually a representation of a tensor called Wilson-Dirac operator. The matrix is  always sparse and structured, so the
iterative methods are definitely considered. Therefore, the procedure
of the application of this operator, resulting in a vector-matrix
product, appears as a critical computation kernel that should be
optimized as much as possible. Due to the size of the matrix, it has to be recalculated on every iteration. Therefore, in the simplest case, evaluating the Wilson-Dirac operator involves each node and 8 neighbors. Such
configuration is really hindering in terms of computation, as memory access is very far from sequential and standard methods fail miserably.  For current and future generation of
supercomputers the hierarchical memory structure makes it next to
impossible for a physicist to write an efficient code. 

But even for computer scientists, the rapid change in parallel architectures makes the design of an
optimized LQCD simulation a real challenge. 
This requires to design, select and combine
iterative methods and preconditioners adapted to the problem and the
target architecture, to optimize data layout and organize parallelism
between nodes, cores, accelerators and SIMD units. In order to
harness all resources of the hardware, orchestrating the work on many
cores and accelerators, using different levels of parallelism, complex
memory hierarchies and interconnect networks takes a large part of the
tuning time, often at the expense of the exploration of new
algorithms/preconditioners. Indeed, testing new methods can only be
achieved with large enough data sets, requiring efficient parallel
codes. Several codes and libraries have been designed for Lattice QCD
 and many works have been published on code optimization for
Lattice QCD, among them the studies on Blue Gene/Q~\cite{Doi:2012},
Intel Xeon Phi~\cite{qcdphi} or clusters of
GPUs~\cite{Joo:2012:LQG:2403996.2404003}.  They offer some degree of flexibility, but usually only focus at a small subset of existing architectures.
However, designing new
iterative methods, combining existing ones, changing data layout
within these frameworks and tools is difficult and requires a
significant code rewriting effort. This clearly hinders the adaptation
of code to the new parallel machines, limiting performance and the
expected scientific results.

This paper proposes a domain-specific language (DSL), QIRAL, for the
description of Lattice QCD simulations, and its compiler to generate
parallel code.  A DSL can help to separate the high level aspects of the simulation
from machine-dependent issues. The contribution of QIRAL is to address
this twofold challenge:
\begin{itemize}
\item Propose to physicists a domain-specific language expressive
  enough to enable the description of different models and algorithms,
  and more importantly, expressive enough to enable algorithmic
  exploration by composing different algorithms and preconditioners
  as well as the design of new algorithms.
\item Generate from this description efficient codes for parallel
  machines. Explicit parallelism and data layout are automatically generated and can be guided by the user. The code generated by QIRAL targets shared
  memory parallel machines, corresponding to one node of larger Lattice QCD
  simulations. This code uses OpenMP and a library
  for efficient SIMD operations.
\end{itemize}
With a higher level description of the Lattice QCD formulation we achieve multiple goals. It
becomes easier to try new algorithmic ideas, the high level code is
easier to maintain and develop. This  makes numerical simulation
accessible to a large number of users, not necessarily high performance
computing experts.  We show on several architectures, from Nehalem-EX with 128
cores to the Xeon Phi accelerator that the code generated with QIRAL
competes in terms of parallel efficiency and performance with tmLQCD,
while QIRAL provides an easier framework for the writing of algorithms
and the adaptation to new architectures.

This paper is organized as follows: first we describe the DSL in
Section~\ref{sec:dsl}, describe the high-level compiler in
Section~\ref{sec:implem}. Then the optimizations for locality, parallelism and
SIMDization are presented in Section~\ref{sec:opt}. Benchmarks, comparing with tmLQCD and describing strong scalability are
shown in Section~\ref{sec:perf}.

The whole project, under the name PetaQCD~\cite{ANGLESDAURIAC:2010:IN2P3-00380246:1}, 
was partly funded by a grant from ANR, 
through the program COSINUS-2008, from 2009 up to 2011.

\section{The QIRAL Domain-Specific Language}
\label{sec:dsl}
As one of the purposes of the QIRAL DSL is to give scientists a
familiar tool to describe the problem in scope, it makes sense to take
an existing system of symbolic notation as the basic language. There
are two such systems in most disciplines, \LaTeX{} and
Mathematica. While we are not attached to a particular one, we chose
to use \LaTeX{}-like syntax where certain additional macros have been
defined. Therefore the QIRAL description can be processed either using
the QIRAL compiler to produce program source code or alternatively, by
the \LaTeX{} typesetter to produce its documentation. This means we revive the principle of
\emph{literate programming} coined by Donald
Knuth~\cite{literateprogramming}. For instance, the algorithm in
Figure~\ref{fig:CG} is a QIRAL program included into this document as
processed by \LaTeX{}. The description of the language given in the following complements a
description previously presented by the authors~\cite{places}.

QIRAL is a language for describing linear algebra objects and operations, with a specialization in 
manipulation of sparse matrices defined through tensor products and
direct sums of dense matrices as they appear in Lattice QCD. It is well-adapted to the particular features of Lattice QCD, or, rather, any realistic field theory. All interactions in such theories are local, therefor it is well-suited for the so-called stensil computation, involving a point and its nearest neighbours only. It assumes that the underlying manifold is a 4D space-time, cut into a cartesian mesh (the lattice).
 The Dirac operator, used for this
inversion, is a sparse but regularly structured matrix that can be
seen as a diagonal of dense matrices. This operator describes interaction between matter fields and guides the evolution of the system.
The matter fields are represented by spinors, which are complex matrices having four spin  and three color. Hence for a $24^3\times 48$ lattice, the Dirac operator is a
matrix of $(24^3\times 48 \times 12)^2$ complex values. Due to locality of the interaction, the resulting matrix is sparce, and QIRAL takes advantge of this structure to reduce the unnecessary computation.

Elements of the language are declarations, equations, algorithms and the
goal. Declarations declare symbols and functions with their
type. Basic types are boolean, integers, real ($\Real$), complex
($\Complex$) vectors (\texttt{V}), matrices (\texttt{M}), indices and
index sets. Vectors and matrices are defined over index sets either
defined through the notation \texttt{V1}$[is]$, where $is$ is the
possibly multi-dimensional index set for vector \texttt{V1}, or
deduced through type inference. A particular element of a vector is
accessed by the use of an index:
\texttt{V1[I1]}. Figure~\ref{fig:dirac} shows the declaration of the
constants used for Lattice QCD, and the definition of Dirac operator as a matrix. The two other matrices, $\Pe$ and $\Po$ are projections, keeping only black or white elements of the lattice, like a 4D checkerboard. 

Equations are used to define variables or functions. 
Figure~\ref{fig:id} describes nearly all properties and definitions on the constant and functions used for the simulation. For instance, 
the function ``invertible'' is defined  only for some expressions. 

\noindent\begin{minipage}[b]{0.46\linewidth}\centering

{\small
\textbf{Constant:}
\begin{align*}
\Dirac, \Pe, \Po, \gamma_5 & \in M, \\
L, S, C, even & \in Indexset, \\
\gamma& \in Index -> M,\\
\U & \in Index -> M , \\
\kappa, \mu, & \in \Real,\\
D & \in Indexset
\end{align*}
\textbf{Variable:} $s \in Index,  d \in Index$

\begin{align*}
\Dirac
  & = \id{L \tensor C \tensor S}   \\
  & + 2 * i * \kappa * \mu * \id{L \tensor C} \tensor \gamma_5 \\
  & + - \kappa * \ssum{d \in  D}{((\shift{L}{-d} \tensor \id{C}) * \dsum{s \in L}{\U[s \tensor d]}) \tensor ( \id{S}  - \gamma[d])} \\
  & + - \kappa * \ssum{d \in  D}{((\shift{L}{d} \tensor \id{C}) * \dsum{s\in L}{\U[s \tensor - d]}) \tensor ( \id{S}  + \gamma[d])} \;\\
  \Pe &= \pr{even}{L} \tensor \id{C \tensor S} \;\\
  \Po &= \pr{!even}{L} \tensor \id{C \tensor S}\;
\end{align*}
}
\captionof{figure}{\small Definitions of the Dirac matrix on a Lattice $L$ in QIRAL, and the two projections for even and odd elements ($\Pe$ and $\Po$ respectively) of this lattice.}
\label{fig:dirac}
\end{minipage}%
\begin{minipage}[b]{0.46\linewidth}\centering
{\small
\textbf{Constant:}
\[dx, dy, dz, dt \in Index \]\noindent
\begin{align*}
D & = \{ dx, dy, dz, dt\} \; \\
\isPeriodic{L} & = true \; \\
\U[s \tensor d]^\dagger &= \U[(s + d) \tensor - d] \; \\
\U[s \tensor - d]^\dagger &= \U[(s + - d) \tensor d] \; \\
\Preconditionera(Dirac) & = Pe \;  \\
\Preconditionerb(Dirac) & = Po \;  \\
\gamma[ d]^\dagger & = \gamma[d] \; \\
\isDiagonal{\gamma_5} & = true \; \\
\gamma_5 * \gamma_5 & = \id{S} \; \\
\gamma_5 * \gamma[d] & = - \gamma[d] * \gamma_5 \; \\
\isInvertible{\id{S} + c * \gamma_5} & = true \; \\
\isInvertible{\id{S} - c * \gamma_5} & = true \; \\
\isInvertible{- (c * \id{S}) + i * \gamma_5} & = true \; \\
\gamma_5^\dagger & = \gamma_5 \; \\
\type{\gamma[d]} & = \matrixtype{S}{S} \; \\
\type{\U[s \tensor d]} & = \matrixtype{C}{C} \; \\ 
\type{\gamma_5} & = \matrixtype{S}{S} \;  \\
\vol{S} &= 4 \; \\
\vol{C} &= 3 \; \\
\end{align*}
}
\captionof{figure}{\small Identities of constants used for Lattice QCD.\label{fig:id}}
\end{minipage}

Algorithms are given as possible definitions for statements or expressions. For
instance, the conjugate gradient algorithm in Figure~\ref{fig:CG} provides
the code that computes expressions of the form $x = A^{-1} * b$, when
$A$ and $b$ are given. It outputs the value of $x$, i.e. solves the
linear system $Ax = b$.
\begin{template}{CGNR}
\Input{$A\in M , b \in V$}
\Output{$x \in V$}
\Constant{$\epsilon\in\Real$}
\Match{$x = A^{-1} * b \;$}
\Var{$r, p, Ap, z \in V , \alpha, \beta, n_r, n_z, n_{z1}\in \Real$}
     	$r = b$ \;
	$z = A ^\dagger * r$\; 
	$p = z$ \; 
        $x = 0$ \;
	$n_z = (z \hdp z)$ \;
        $n_r = (r \hdp r)$ \;
	\While{$(n_r > \epsilon)$} {
	  $Ap = A * p $\;
	  $\alpha = n_z / (Ap \hdp Ap)$ \;
	  $x  = x + \alpha * p$ \;
	  $r = r - \alpha * Ap$ \;
	  $z = A ^\dagger * r$ \;
	  $n_{z1} = (z \hdp z)$ \;
	  $\beta = n_{z1} / (n_z)$ \;
	  $p = z + \beta * p$ \;
	  $n_z = n_{z1}$ \;
          $n_r = (r \hdp r)$ \;
	}
\caption{Conjugate Gradient, normal residual method (CGNR).\label{fig:CG}}
\end{template}

\begin{template}{schur}
\Input{$A, P_e, P_o \in M , b \in V $} \Output{$x \in V $}
\Match{$x = A^{-1} * b \;$}
\Constant{$D_{11}, D_{12}, D_{21}, D_{22} \in M$}
\Var{$v_1, v_2, x_1, x_2 \in V$}
\Require{$\isInvertible{P_e * A * P_e ^t}$}
$D_{21} = P_o * A * P_e^t$ \; 
$D_{11} = P_e * A * P_e^t$ \; 
$D_{22} = P_o * A * P_o^t$ \; 
$D_{12} = P_e * A * P_o^t$ \; 
$v_1 = P_e *  b $ \;
$v_2 = P_o * b$ \; 
$x_2 = (D_{22} - D_{21} * D_{11}^{-1} * D_{12})^{-1} * (v_2 - D_{21} * D_{11}^{-1} * v_1)$ \; 
$v_1 = P_e * ( 2 * \kappa * b )$ \;
$x_1 = D_{11}^{-1} * (v_1 - D_{12} * x_2) $ \; 
$x = P_e^t * x_1 + P_o^t * x_2$ \;
\caption{Definition of Schur complement method\label{fig:schur}.}
\end{template}

 The initial statement, in the \textbf{Match} clause, is then defined
 (and replaced) by the pseudo-code. The \textbf{Var} keyword declares
 the type of local variables.  This algorithm is written using the
``algorithm2e'' package in \LaTeX{}, and is not specific to Lattice QCD. 
The user has the possibility to write new algorithms for Lattice QCD or any other algorithm found in common literature. The QIRAL
 compiler finds automatically how to compute for instance $A * p$ when
$A$ is instantiated with the Dirac operator.

Most often the validity of an algorithm depends on prerequisites, special
properties the inputs must have.
These prerequisites are declared in a clause \textbf{Require}
and is proved by the QIRAL compiler. The
following example illustrates this prerequisite mechanism.
Figure~\ref{fig:schur} describes the Schur complement method that is used as
a preconditioner for the conjugate gradient and the conjugate residual. The
condition $\isInvertible{P_e * A * P_e^t}$ is proved automatically by
the compiler when $A$ matches the matrix $Dirac$. To prove this, the
property defined previously for the function ``invertible'' is used. If
the compiler is not able to prove the requirements attached to an
algorithm, the algorithm is not applied and an error is generated.
Notice that on this preconditioning, the statements involve
computation of inverse matrices. For the expression $D_{11}^{-1}$, the
QIRAL compiler can prove automatically that $D_{11}$ is diagonal (when
$A$ is the Dirac operator), and knows how to invert this matrix. For
the computation of the expression $(D_{22} - D_{21} * D_{11}^{-1} *D_{12})^{-1}$, an iterative method has to be applied.

The goal defines the initial code and the list of algorithms to
apply. The algorithms are composed from right to left.  
\begin{goal}{goal}
\Input{$bb\in V$}
\Output{$xx\in V$}
\Templates{CGNR schur} 
$xx[L \tensor C \tensor S] = 
   Dirac^{-1} * bb[L \tensor C \tensor S]$ \;
\end{goal}

For this goal here the preconditioner \texttt{schur} is applied on the initial
statement, and then the CGNR algorithm. It is possible to chain
multiple algorithms, used to apply preconditions before the solvers.
The index set $L\tensor C \tensor S$ represents the Cartesian product
of these sets and the domain for the vector $bb$. At this level, there
is no implicit data layout for vectors and matrices. The vector $bb$
could be either a 4D array of structures, one dimension for each
dimension of L and the structure representing elements indexed by $C$
and $S$, or a 1D array of structures, or just a large 1D array of
complex values. This is orthogonal to the expression of the algorithm.

The output of QIRAL compiler is a function in C and OpenMP pragmas representing the
computation described in the goal, and taking as parameters $bb$ and
$xx$. All other constant values (in particular constant matrices) are
assumed to be global.


\section{Implementation Details}
\label{sec:implem}
The QIRAL compiler is based on a rewriting system, Maude~\cite{maude}. The 
different steps of this transformation are explained
in this section.

\subsection{Algorithms composition and expression simplification}
Algorithms are translated into rules of the rewriting system, while
equations define the equational theory for the rewriting system. The
first step consists in transforming \LaTeX{} input into a Maude
program. Additional modules, defining usual algebraic simplifications
are added to this code. 

The first step parses the \LaTeX{} input and captures only what is described
in predefined environments, for algorithms, definitions and
the goal. Syntactic verification as well as type
checking is performed. The output generated is a Maude module, with equations
corresponding to definitions, rules corresponding to algorithms, and a
unique Maude statement, corresponding to the goal.

The algorithms declared in the goal are applied, in turn, to the
statements provided. These statements are terms for Maude. The
\textbf{Match} clause is the left-hand side (lhs) of the rule, while
the pseudo-code corresponds to a term that is the right-hand side (rhs) of
the rule. Any statement matching the lhs will then be rewritten in the
rhs. If a \textbf{Require} clause exists, it
constitutes the condition for the rewriting. The first statement is
provided by the goal, then algorithms are applied successively.

Definitions and properties are considered by Maude as defining the
equational theory for the rewriting system. Actually, these equations
are handled as automatic rewriting rules: Maude automatically applies
all possible equations, rewriting their lhs into rhs, until the term
is normalized. For instance, the property $x * (y + z) = x * y + x *
z$, stating the distributivity of $+$ over $*$, will only be used to
distribute the operators, not to factorize terms. 

The main objective of this formal rewriting is to eliminate all terms
that are equal to zero. In Lattice QCD, the Dirac matrix used in the
problem is sparse, but built from dense matrices with a regular
structure. To obtain such simplifications, an additional module
defines properties for the linear algebra operators, on complex,
vectors and matrices: Addition is commutative and associative,
multiplication is associative and distributes over the addition,
binary subtraction is converted to unary minus, transposition
distributes over the addition and multiplication, etc. Moreover,
some properties are also defined for permutation and projection
matrices, in particular to handle Schur preconditioning.

\section{Code Optimizations}
\label{sec:opt}

\subsection{Improving Locality}
Loops fusion is a transformation to reduce reuse distances, hence
improving locality. To check if fusion is valid, a simple dependence
analysis, based on dependence distance, is computed. The fusion method
is applied on consecutive independent loops that share the same
iterators, and is applied on all code until no more fusion is
possible. This simple strategy is sufficient for Lattice QCD generated
codes.

Following this fusion, the regions of arrays that are written/read by
all loops, and the regions that are inputs/outputs of loops are
computed. All arrays that are used only in one loop are scalar
promoted. The resulting values are allocated on the stack, and aligned
for further vectorization. This reduces memory consumption.

Both transformations are applied automatically. 

\subsection{Versioning Matrix Multiplication}
The computation involve many multiplications of vectors by constant
matrices, accounting for transformations on spin and color ($S$ and
$C$ index sets respectively). These matrices are small, of size
$3\times 3$ and $4\times 4$ respectively, and the latest have only 2
non-null elements per line, these elements being among
$\{1,-1,i,-i\}$. Therefore specialization of these products is
necessary in order to obtained better performance. These matrix-vector
multiplications appear in expressions of the form 
$(M_1 \tensor M_2) \cdot V$ with $M_1$ and $M_2$ the two matrices multiplied by a tensor
product, and $V$ is a $12$ element vector. In this case the QIRAL compiler uses the
identity
\[(M_1 \tensor M_2) * V = M_1 * V * M_2^t\]
where on the lhs, $V$ is considered a matrix of size
$3\times 4$ and $*$ stands for the matrix product. Therefore, instead of
using general matrix multiplication, QIRAL compiler finds these
occurrences and calls versioned matrix multiplications.  Specializing
such multiplications for these particular sizes, in particular
performing SIMDization, is essential for performance. These functions
correspond to the hot-spot of the codes generated by QIRAL.  The codes
of these functions are hand-written in \texttt{libqiral} library as
presented in Figure~\ref{fig:compiler}.

Other expressions can be replaced by library calls, and QIRAL
changes expressions on vectors and matrices into BLAS calls (or
specialized BLAS).  The fact that the QIRAL compiler automatically
identifies these functions in the code generated from the different
algorithms facilitates the optimization process and is
an asset of QIRAL. The optimization of these functions in
\texttt{libqiral}, specializations of BLAS, can indeed be achieved by
an expert in high-performance computing, independent of any Lattice QCD context.

\section{Performance Results}
\label{sec:perf}

\begin{figure}[ht]
\includegraphics[width=0.45\textwidth]{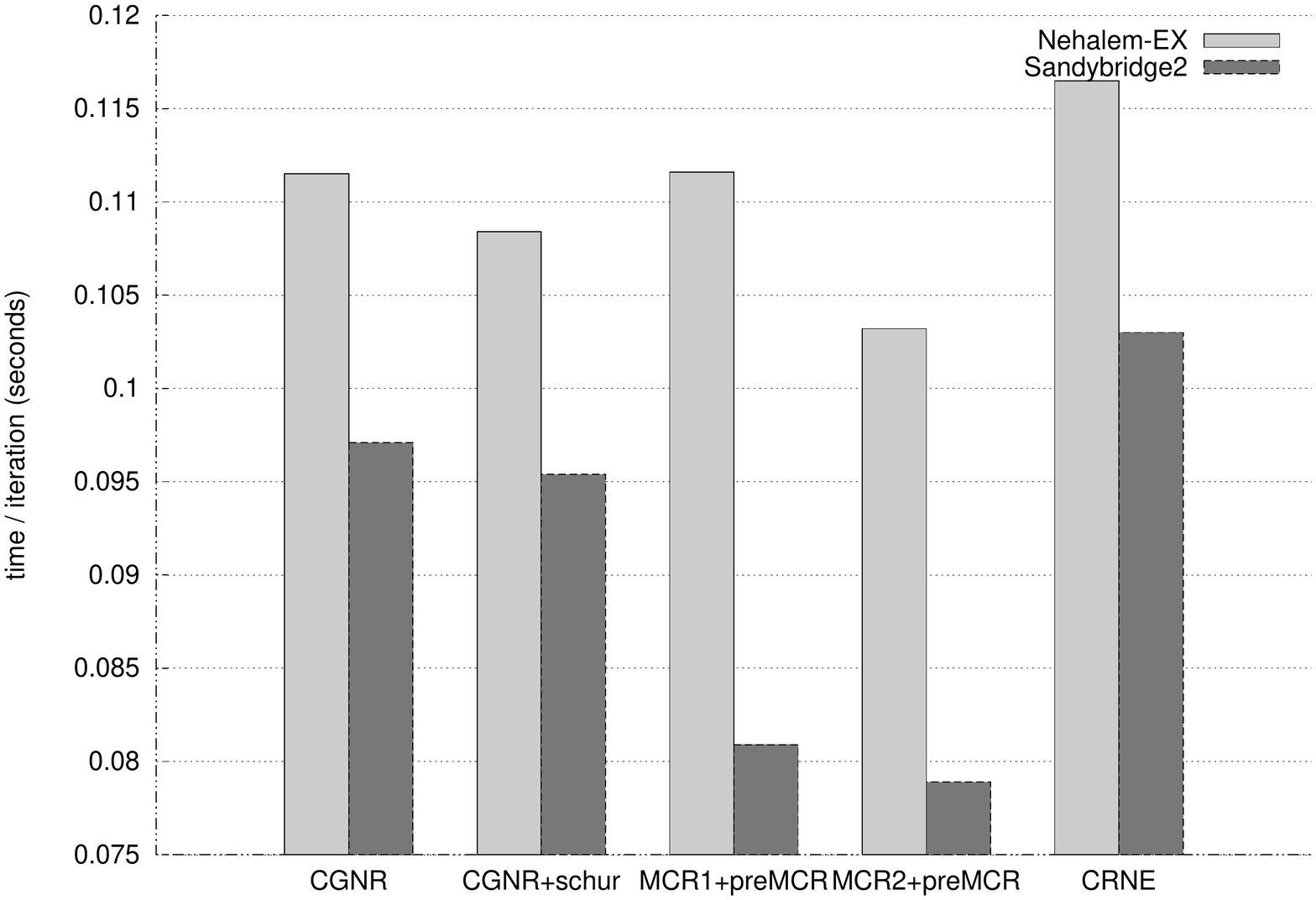}\vspace{-1cm}
\includegraphics[width=0.45\textwidth]{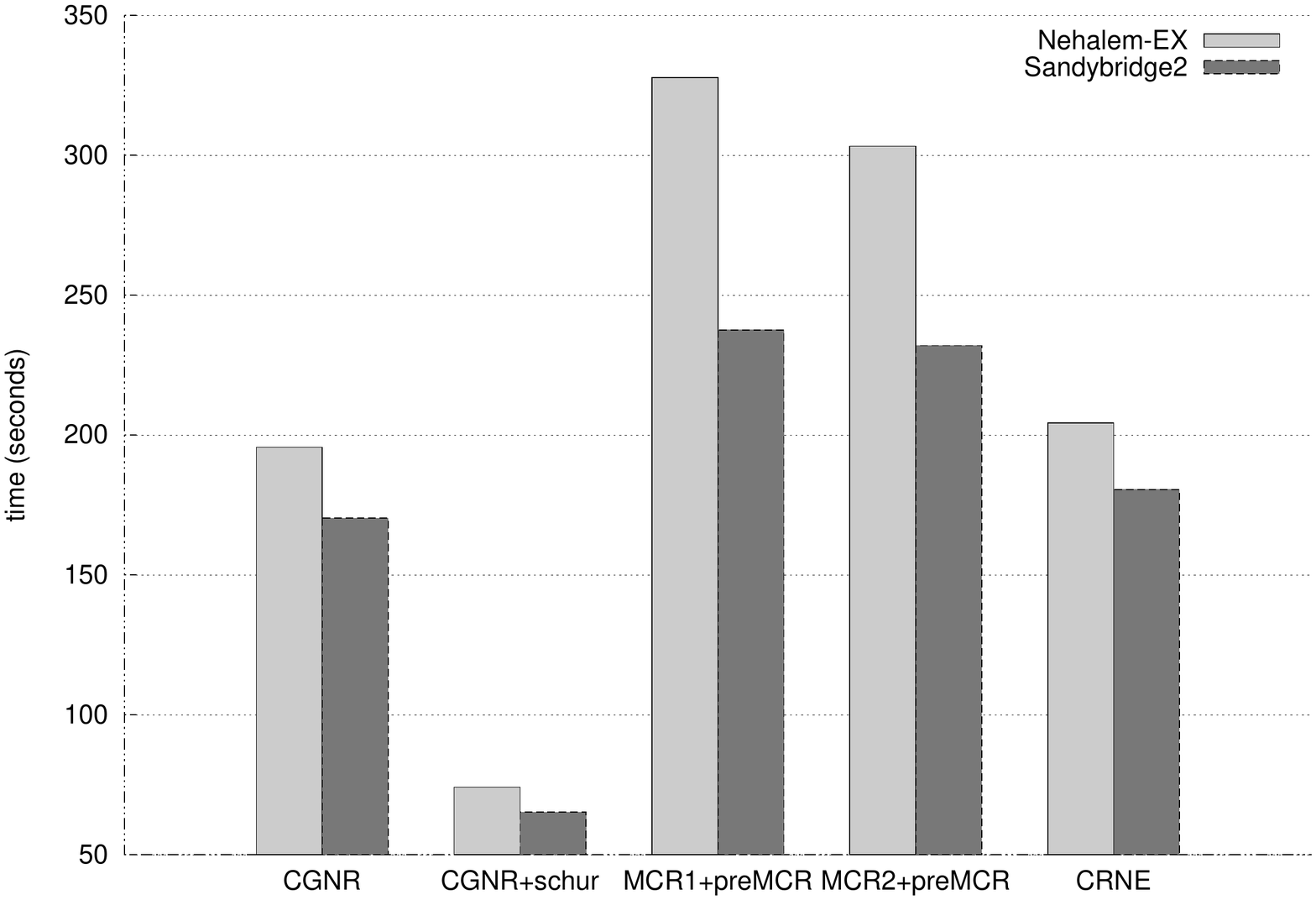}
\caption{Comparison between different iterative methods, on Nehalem-EX
  and Sandybridge 2 architectures. Left figure:  Time in seconds per iteration. Right figure:  Total execution time. \label{fig:algosxp}}
\end{figure}

Several iterative methods  are written with QIRAL. Figure~\ref{fig:algosxp} presents some of these methods, for two architectures: CGNR, CRNE, MCR1 and MCR2 with some preconditioners: Schur and preMCR. 
We observe that while MCR2 exhibits the best time per iteration, the
method takes more time to converge than CGNR and Schur. This shows
that the best method cannot be determined only by benchmarking a
single iteration, but it is necessary to run all iterations.  Besides, the second plot of Figure~\ref{fig:algosxp} 
shows that the relative difference may vary
according to the architecture. While the absolute best method is still
the same (here CGNR combined with Schur), this stresses the fact that
the algorithmic solution may be chosen depending on the target
architecture.

In order to compare tmLQCD with the code generated by QIRAL, the same
algorithm is used for both (CGNR and Schur
preconditioning). Performance is displayed for all architectures as the
total execution time multiplied by the number of cores.  Due to the
fact that tmLQCD is using MPI, there is no version for Xeon Phi.
Besides, the tmLQCD code uses in-line assembly code with SSE3
instructions. Adapting this code for newer SIMD extensions is more
difficult than adapting intrinsics as used by QIRAL. Indeed for
intrinsics, part of the optimization work still relies on the
compiler: register allocation, generation of FMAs, scheduling.  The
code generated by QIRAL has been quickly ported to these
architectures, and then code tuning has focused on the library used by
QIRAL (with versioned BLAS), using intrinsics and aggressive in-lining.

\begin{figure}[ht]
\includegraphics[width=.45\textwidth]{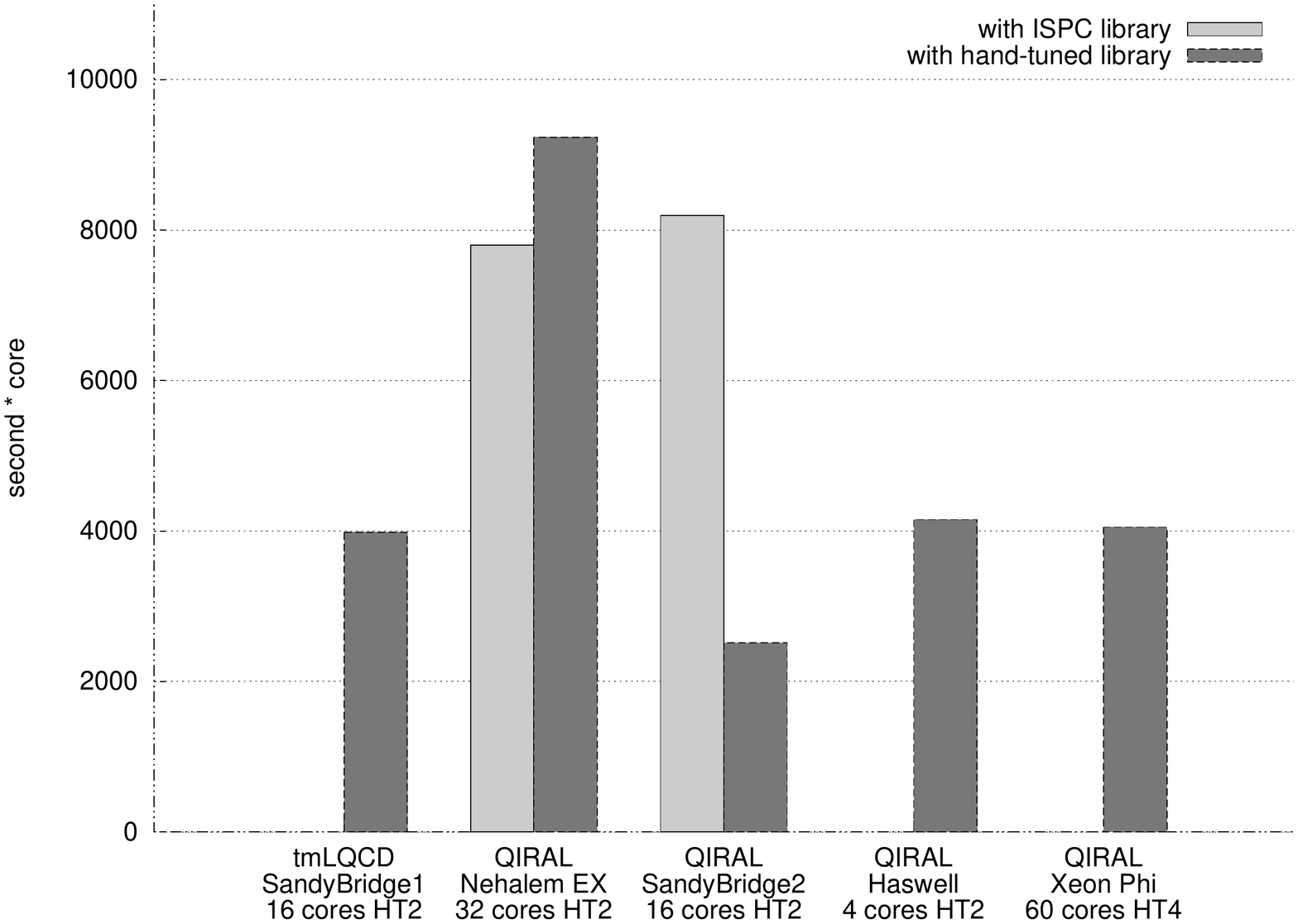}
\includegraphics[width=0.45\textwidth]{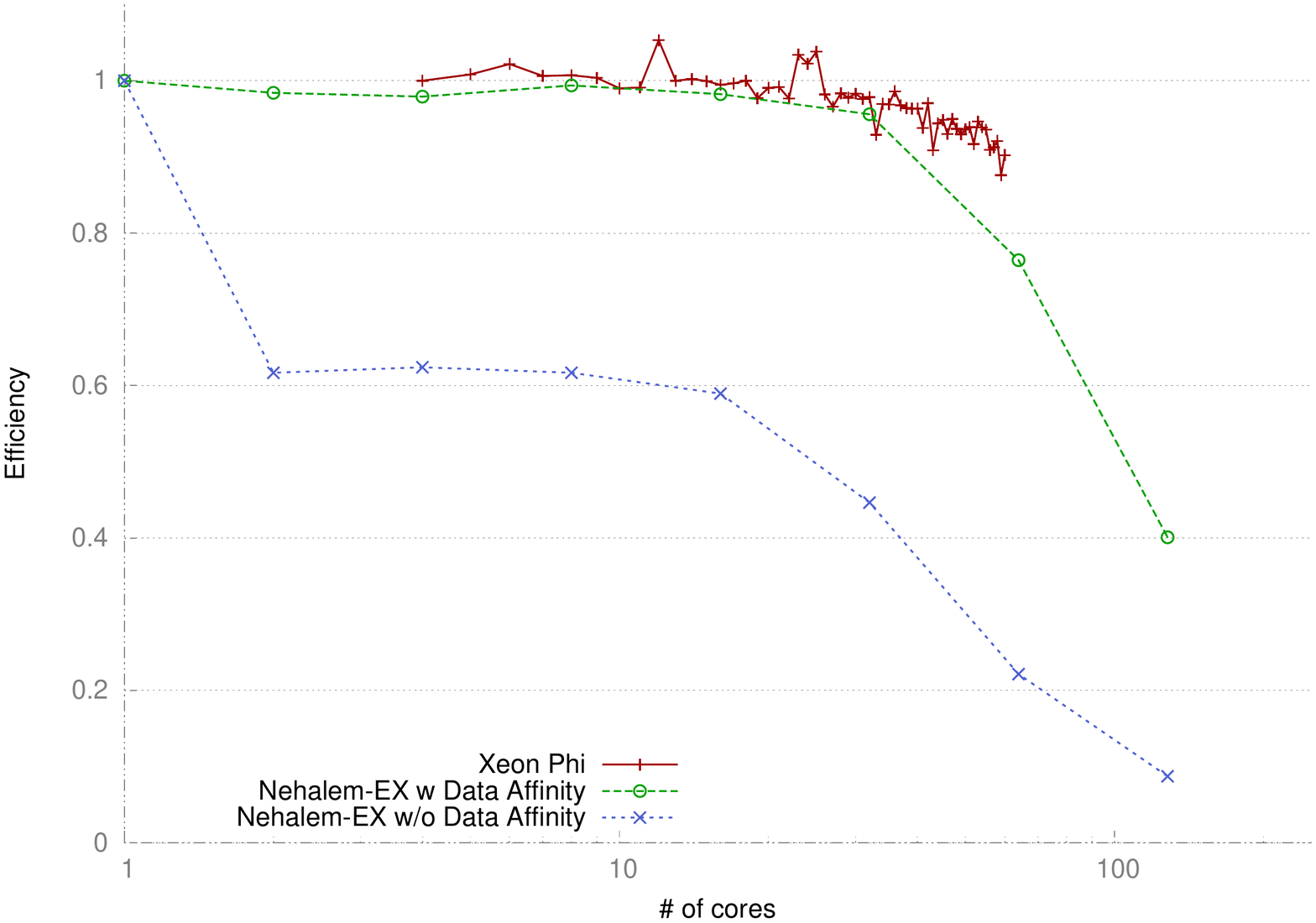}
\caption{Left: Normalized performance for the inversion on different
  architectures, with QIRAL and tmLQCD codes. Performance is shown in
  sec*core (ie ``seconds times number of cores''), 
  lower is better.  The execution time is obtained by
  dividing this performance by the number of cores. The same method, a
  conjugate gradient with Schur preconditioning is used in all
  cases, with a lattice of size $24^3*48$ and an error of
  $10^{-14}$.\label{fig:cgnr} Right: Efficiency of the code generated by QIRAL on Xeon Phi and
  Nehalem EX, according to the number of cores used. For both
  architectures, the method used is the conjugate gradient with
  Schur preconditioning. The lattice size for the Xeon Phi is
  $24^3*48$ and for the Nehalem-EX, $64^3*128$. On the Nehalem-EX the
  efficiency is measured with and without NUMA-aware memory
  allocation.}
\end{figure}

Figure~\ref{fig:cgnr} presents timing results on different
architectures, comparing tmLQCD code with QIRAL generated code.  For
QIRAL, the ``hand-optimized library'' corresponds to the best version
obtained, using intrinsics (AVX, AVX2, Xeon Phi) for Sandybridge,
Haswell and Xeon Phi architectures. The Nehalem EX version does not
use SSE SIMD intrinsics. This explains why QIRAL/Nehalem EX version is
more than two times slower than tmLQCD. For Xeon Phi, the performance
displayed corresponds to the use of all the $60$ cores, and a linear
speed-up can be observed by using an increasing number of cores. The
ISPC compiler has been used to generate SIMD version of matrix
multiplication of size $3\times 4$ on complexes. The compiler is still under heavy
development and does not fully work for Xeon Phi. Figure~\ref{fig:cgnr}
shows that the level of performance reached with ISPC is not competing
with the level for hand-tuned intrinsics.

The strong scalability of the code generated by QIRAL is evaluated on
Xeon Phi and Nehalem-EX architectures. Figure~\ref{fig:cgnr}, right, shows
efficiency results for different number
of cores. Note that the size of the lattice is different for both
architectures, reflecting the need for different granularity. The
efficiency for the Xeon Phi is compared to the run on 4 cores, with 4
threads each. This explains why for some number of threads, the
efficiency goes beyond 1. The code scales well up to the $60$ cores
($240$ threads). For the Nehalem-EX machine, the efficiency is higher
than $95\%$ up to $32$ cores, and then drops quickly. The reason is
that a 128-core node is structured with 4 groups of 4 octo-cores,
connected through a switch. Going through the switch has a high
penalty in terms of performance.

\section{Conclusion}
The contribution of this paper is a new domain-specific language,
QIRAL, for the automatic code generation of OpenMP codes for Lattice
QCD simulations. QIRAL language offers to physicists the possibility to
implement iterative methods and preconditioners, literally ``from the
book'' using \LaTeX{}, or design new ones, and test them on large
parallel shared memory machines or on accelerators such as the Xeon
Phi. The language enables the composition of preconditioners and
iterative methods, and the compiler checks automatically the validity
of application for each method. This makes possible a more systematic
exploration of the algorithmic space:
 indeed, it removes from the physicists the burden of long and stressful 
validations of their new code since it will be automatically generated, 
then safer, and the time-to-market for a viable product will be much shorter.
 The QIRAL compiler generates
OpenMP parallel code using BLAS or specialized versions of BLAS
functions. Further hand-tuning is possible on the code generated by
QIRAL, and we have shown that the performance on various multi-core
architectures and on Xeon Phi accelerator it compares or outperforms
the performance of a hand-tuned Lattice QCD application, tmLQCD.

Among the perspectives of this work, the automatic generation of a
communication code for multi-node computation would enable to run
Lattice QCD simulations on a larger scale. Besides, the fine tuning of
the library functions used by QIRAL on different architectures, in
particular their SIMDization, could be improved.

\section{The bibliography}

\bibliographystyle{abbrv}
\bibliography{qiral}

\end{document}